\documentclass[twoside,final]{pro12}

\usepackage[ansinew]{inputenc}
\usepackage{amsmath}
\usepackage{graphicx} 

\usepackage{color}

\head{G. Hodos\'an et al.} {Radio emission of lightning on HAT-P-11b}	

\begin{document}

\title{EXO--LIGHTNING RADIO EMISSION: THE CASE STUDY OF HAT-P-11b}

\author{G. Hodos\'an\adress{\textsl SUPA, School of Physics and Astronomy, University of St Andrews, St Andrews KY16 9SS, UK; hodosan.gabriella@gmail.com}$\,\,\,$\adress{\textsl Centre for Exoplanet Science, University of St Andrews, St Andrews, UK}$\,\,\,$\adress{\textsl Instituto de Astrof\'isica de Andaluc\'ia, (IAA-CSIC), Glorieta de la Astronom\'ia s/n, 18008, Granada, Spain}$\,\,$, Ch. Helling$^{* \, \dagger}$\adress{\textsl Anton Pannekoek Institute for Astronomy, University of Amsterdam, Science Park 904, 1098 XH Amsterdam, the Netherlands}$\,\,$, and P. B. Rimmer\adress{\textsl Astrophysics Group, Cavendish Laboratory, J.J. Thomson Avenue, Cambridge CB3 0HE, UK}$\,\,\,$\adress{\textsl MRC Laboratory of Molecular Biology, Francis Crick Avenue, Cambridge Biomedical Campus, Cambridge CB2 0QH, UK}}



\maketitle

\begin{abstract}
Lightning induced radio emission has been observed on solar system planets. Lecavelier des Etangs et al. [2013] carried out radio transit observations of the exoplanet HAT-P-11b, and suggested a tentative detection of a radio signal. Here, we explore the possibility of the radio emission having been produced by lightning activity on the exoplanet, following and expanding the work of Hodos\'an et al. [2016a]. After a summary of our previous work [Hodos\'an et al. 2016a], we extend it with a parameter study. The lightning activity of the hypothetical storm is largely dependent on the radio spectral roll-off, $n$, and the flash duration, $\tau_\mathrm{fl}$. The best--case scenario would require a flash density of the same order of magnitude as can be found during volcanic eruptions on Earth. On average, $3.8 \times 10^6$ times larger flash densities than the Earth--storms with the largest lightning activity is needed to produce the observed signal from HAT-P-11b. Combined with the results of Hodos\'an et al. [2016a] regarding the chemical effects of planet-wide thunderstorms, we conclude that future radio and infrared observations may lead to lightning detection on planets outside the solar system.
\end{abstract}

\section{Introduction}

\begin{figure}[ht]
\centering
\includegraphics[width=0.8\textwidth]{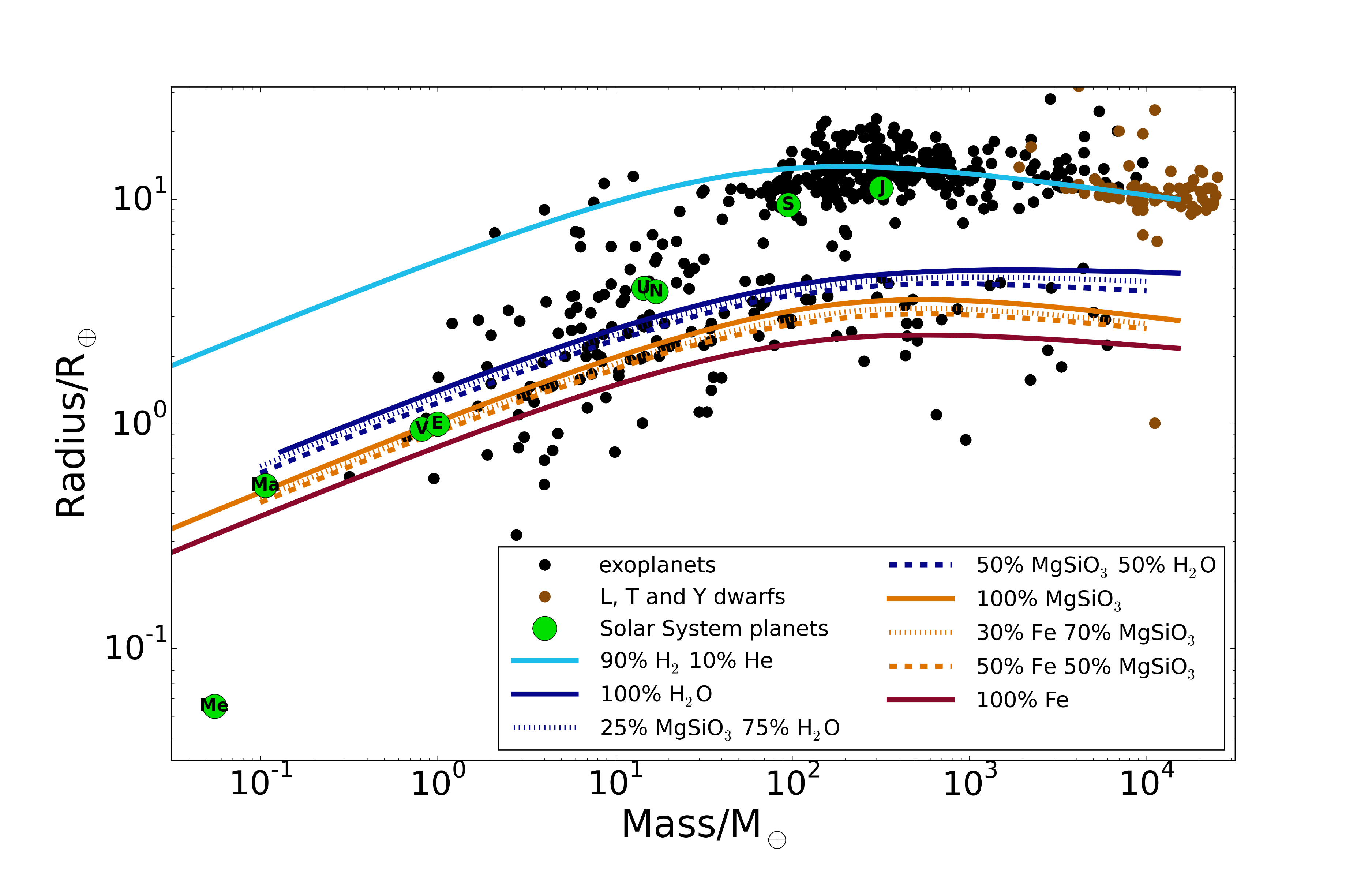}
\caption{Diversity of exoplanets (black dots) and brown dwarfs (brown dots) on a radius--mass plot (in Earth values, M$_{\oplus}$, R$_{\oplus}$). The coloured lines show mass--radius relationships for various bulk compositions (see Hodos\'an et al. [2016a]). We note that for some cases the uncertainties in mass and radius are large enough to move the planet from one compositional region to the other. The solar system planets are also shown (green circles) for comparison. The large diversity of planets in size, mass, and chemical composition, result in a large variety of surface and atmospheric environments. Source of data: exoplanet.eu}
\label{fig:1}
\end{figure}

Since the discovery of the first exoplanet around a Sun--like star [Mayor and Queloz, 1995], more than 3500 planets orbiting other stars have been confirmed\footnote{exoplanet.eu, 2017. Feb. 22.} to exist. This large number provides us with an adequate sample for statistical studies, and helps us learn more about the place of the solar system in the multitude of extrasolar systems. We can focus on the more detailed characterization of the different types of planets, including atmospheric chemistry and internal composition. Figure~\ref{fig:1} illustrates the large variety of exoplanets observed so far, including hot Jupiters, mini--Neptunes (e.g. HAT-P-11b), super--Earths, and even planets smaller than Earth. The diversity of exoplanets in mass and radius result in divers bulk compositions, which, together with a wide range of distances from the host star, will suggest that these objects host a large variety of surface and atmospheric environments. Observational campaigns have revealed that lightning occurs in very diverse environments on Earth and on other solar system planets. Lightning is frequently produced in thunderclouds that are made of water and ice particles, and occasionally in clouds of ice and snow particles, producing "winter lightning" [e.g. Brook et al., 1982]. During explosive volcanic eruptions intense lightning activity is observed in volcano plumes, which are primarily composed of mineral dust particles [James et al., 2008, and references therein]. Sand storms, dust storms and dust devils are also well-known hosts of electrical activity [e.g. Yair et al., 2016]. On Jupiter and Saturn, lightning is produced in dense, vertically extended, convective clouds [e.g. Dyudina et al., 2004; 2013]. Therefore, the large variety of lightning hosting environments and extrasolar bodies suggests that lightning occurs outside the solar system [Helling et al., 2016a].

Lightning indicates cloud formation and large-scale convection in atmospheres of planetary objects. It supports the global electric circuit (GEC) [Wilson, 1921], a continuously present electric current between the earth--ionosphere cavity. Furthermore, the Miller--Urey experiment showed that in highly reductive environments (composed of CH$_4$, NH$_3$, H$_2$, and H$_2$0), such as that thought to be found in early Earth, lightning discharges produce prebiotic molecules [Miller and Urey, 1959], which are important for the formation of life. Though, it is debatable whether the early atmosphere of the Earth was such a reductive one [e.g. Kasting, 1993], the large variety of extrasolar planets suggests that such environments may exist outside the solar system. It has also been shown that for a large variety of prebiotic scenarios [e.g. Patel et al., 2015], hydrogen cyanide (HCN) is the key simple precursor molecular species for pre-biotic chemistry. Ardaseva et al. [2017] found that lightning is very effective at generating HCN in the atmosphere of the Early Earth (as defined by Kasting [1993]), further supporting the theory that lightning is an important element for prebiotic chemistry. 

Lightning signatures span the whole electromagnetic spectrum from extremely low frequency radio emission to high-energy X-rays [Rakov and Uman, 2003]. When electric current is generated by accelerating electrons in a conducting channel, such as it occurs during lightning, the channel will act as an antenna converting the electric power resulting from the time-dependent current into radio waves [e.g. Zarka et al., 2004]. Lightning radio spectrum covers the few Hz $-$ $10^2$~MHz range [Rakov and Uman, 2003]. Lightning radio emission has been observed not only on Earth, but on Jupiter, Saturn, Uranus, Neptune, and potentially on Venus [e.g. Zarka and Pedersen, 1986; Gurnett et al., 1990; Rinnert et al., 1998; Fischer et al., 2006a; Russell et al., 2008; Yair, 2012]. Recently, radio observations have opened new paths to study properties of extrasolar objects, such as brown dwarfs [e.g. Williams and Berger, 2015], which are only a step away from giant gas planet detections in the radio wavelengths. Electron cyclotron maser emission has been suggested to be one possible mechanism for this observed radio emission [Grie{\ss}meier et al., 2007]. Several campaigns to observe radio emission from extrasolar planets have been started [e.g. Lecavelier des Etangs et al., 2013, and references therein]; however, no radio emission from exoplanets have been conclusively detected. 

In this paper, we summarize our work of possible lightning occurrence on an exoplanet based on previous radio observations of the object [Hodos\'an et al., 2016a]. We focus on the tentative radio emission potentially  detected from HAT-P-11b and how that gives information on a hypothetical thunderstorm in the atmosphere of the planet. Section~\ref{sec:hatp11} summarizes the radio observations of Lecavelier des Etangs et al. [2013]. Section~\ref{sec:radio} introduces a simple model of lightning radio emission, and the important formulas needed to infer lightning activity from its radio signal. In section~\ref{sec:disc} we summarize and discuss our result. In section~\ref{sec:conc} we conclude the paper.

\section{HAT-P-11b: radio observations} \label{sec:hatp11}
The exoplanet HAT-P-11b is a mini--Neptune, with size R$_p$ = 4.7 R$_{\oplus}$ and mass M$_p$ = 26~M$_{\oplus}$ (units of Earth radius and mass) [Bakos et al., 2010]. It orbits its host star, a K4 dwarf, at a distance of $\sim 0.053$ AU [Lopez and Fortney, 2014]. In 2009, Lecavelier des Etangs et al. [2013] (hereafter L13)  observed a tentative radio signal from HAT-P-11b at 150~MHz with the Giant Meterwave Radio Telescope, with an average flux of 3.87~mJy. They conducted the observations during the planetary occultation, when it passed behind the host star, and found that the signal vanished when the planet was not visible. They re-observed the planet with the same instruments in 2010, but did not detect any signal. Assuming that the 150~MHz signal from 2009 was real and was coming from the exoplanet, the non-detection in 2010 suggests that it was produced by a transient phenomenon. L13 suggested that the radio signal was the result of interactions between the magnetic field of the planet and stellar coronal mass ejections or stellar magnetic field. If the mJy radio emission were caused by cyclotron maser emission, a large planetary magnetic field of 50~G would be required [L13]. The strength of the surface magnetic field of the solar system planets ranges between $\sim10^{-4}$~G (Mars) and $\sim 4$~G (Jupiter) [Russell, 1993].
  
The activity of the star may support lightning activity in close-in planets, as the ionizing effects of the star on the planetary atmosphere will be enhanced the closer the planet is to the star [Longstaff et al., 2017]. HAT-P-11b is much closer to its host star than the solar system planets with lightning, resulting in a stronger irradiation from the star. Studies suggest that highly irradiated atmospheres will form clouds in very dynamic environments [e.g. Helling et al., 2016b], which may host high lightning activity [e.g. Helling et al., 2013]. HAT-P-11b, orbiting its host star closely, likely has a dynamic atmosphere that will impact the observable cloud distribution [e.g. Fraine et al., 2014; Line and Parmentier, 2016]. The resulting patchy clouds [Line and Parmentier, 2016] could focus potential lightning activity to a certain region, at certain times covering a large fraction of the planet. These potentially large, dynamical cloud systems could support the occurrence of high lightning rates. Here, we hypothesise that the radio signal observed from HAT-P-11b by L13 was caused by a lightning storm and estimate how much lightning would be needed to produce an average 3.87~mJy radio signal at 150~MHz. We expand our previous study [Hodos\'an et al., 2016a] with the analysis of the effects of various lightning properties on the final results. This parameter study puts our conclusions (section~\ref{sec:conc}) into a new context compared to Hodos\'an et al. [2016a].

\section{Lightning radio emission and flash density model} \label{sec:radio}
We assume that the radio signal measured by L13 was real and was originating from HAT-P-11b. Because of the transient nature of the signal, and because it does not show clear polarization (L13, their Fig. 2), we assume that it was produced by a thunderstorm that was present over the observed disk of the planet, continuously producing lightning flashes throughout the observations. As a first step, we treat lightning on HAT-P-11b with the same physical properties that were measured for Earth or Saturn. Our goal is to determine how much lightning would be required to produce the observed radio flux.

First, we determine the radio power spectral density, $P/\Delta f$ [W Hz$^{-1}$], radiated by one lightning flash at frequency $f$ [Hz]: 

\begin{equation}
\label{eq:1}
\dfrac{P}{\Delta f} = \dfrac{P_0}{\Delta f} \Bigg(\dfrac{f_0}{f}\Bigg)^{\!\! n},
\end{equation}

\noindent where $P_0/\Delta f$ [W Hz$^{-1}$] is the peak power spectral density at peak frequency $f_0$ [Hz], and $n$ is the spectral roll-off at higher frequencies [Farrell et al., 2007]. The radio flux of a lightning flash, $I_{\mathrm{\nu, fl}}$ [Jy], from distance $d$ [m] is obtained from Eq. \ref{eq:2b} through $P/\Delta f$: 

\begin{equation}
\label{eq:2b}
I_{\mathrm{\nu, fl}} = \frac{(P/\Delta f)}{4 \pi d^2} \times 10^{26},
\end{equation}

\noindent where 1~W~Hz$^{-1}$~m$^{-2} = 10^{26}$~Jy. The observed radio flux, $I_{\mathrm{\nu, obs}}$ [Jy], will be the contribution of all the lightning flashes occurring during the observation:

\begin{equation}
\label{eq:2}
I_{\mathrm{\nu, obs}} = I_{\mathrm{\nu, fl}} \frac{\tau_{\mathrm{fl}}}{\tau_{\mathrm{obs}}} n_{\mathrm{tot,fl}},
\end{equation}

\noindent where $\tau_{\mathrm{fl}}$ [h] is the characteristic duration of the lightning flash, $\tau_{\mathrm{obs}}$ [h] is the observation time, and $n_{\mathrm{tot,fl}}$ is the total number of flashes. A lightning flash has a much shorter duration than the observation time, therefore it cannot be considered as a continuous source. As a result, the contribution of one lightning flash to the observed radio flux has to be weighted by its duration time over the observation time. 

We are interested in the number of lightning flashes producing an average 3.87~mJy radio flux, which was observed from the direction of HAT-P-11b. This is given by $n_{\mathrm{tot,fl}}$, which we obtain from Eq. \ref{eq:2}. Then, we convert the result into flash density, $\rho_{\mathrm{fl}}$ [flashes km$^{-2}$ h$^{-1}$], which is compared to lightning occurrence rate observed in the solar system:

\begin{equation}
\label{eq:2a}
\rho_{\mathrm{fl}} = \frac{n_{\mathrm{tot, fl}}}{2\pi R_p^2 \tau_{\mathrm{obs}}},
\end{equation}

\noindent where $R_p$ [km] is the radius of the planet. $\rho_{\mathrm{fl}}$ carries a statistical information on the occurrence of lightning in space [km] and time [h].

\section{Results and Discussion} \label{sec:disc}

We conduct a parameter study to investigate the effects of the spectral roll-off, $n$, and the flash duration, $\tau_\mathrm{fl}$, on the resulting lightning flash densities, $\rho_\mathrm{fl}$. We note that on Earth the spectra roll-off is not constant throughout large frequency ranges. It changes from $f^{-2}$ to $f^{-4}$ [Rakov and Uman, 2003]. This, however, is not implemented in our study, since there is no way of telling where the roll-off would change and by how much in different atmospheres, just like we cannot tell what the roll-off of Saturnian lightning spectra is at frequencies above the limit of Voyager 1 ($\sim 40$ MHz) [Zarka and Pedersen, 1983]. In order to demonstrate the order of thoughts, first we calculate $\rho_\mathrm{fl}$ using a selected set of input parameters described below, and discussed in more details in Hodos\'an et al. [2016a].

First, using Eq. \ref{eq:1}, we determine the power spectral density, $P/\Delta f$, of a lightning flash at $f$ = 150~MHz, the centre frequency of the band that was used to observe HAT-P-11b [L13]. Equation \ref{eq:1} comprises of several unknowns, which we determine assuming that lightning on HAT-P-11b has the same energetic properties as lightning on Saturn. Because the frequency dependence of the power spectral density of lightning is a negative power law (Eq. \ref{eq:1}), the peak power spectral density, $P_0/\Delta f$, and the peak frequency, $f_0$, can be substituted with values of radiated power spectral density, $P_\mathrm{S}/\Delta f$, and measured frequency, $f_S$, on Saturn. Cassini--RPWS measured the radiated power spectral density of SEDs (Saturnian Electrostatic Discharges) to be $P_\mathrm{S}/\Delta f =$ 50~W~Hz$^{-1}$ at $f_\mathrm{S} =$ 10~MHz [Fischer et al., 2006b]. The spectral roll-off, $n$, is determined by the duration of a stroke, the fine structure of a lightning flash [Bruce and Golde, 1941]. A cloud-to-ground lightning flash on Earth is a combination of strokes and interstroke intervals. In our study, we consider the spectral behaviour of an SED that was measured over the time-scale of flashes ($\sim$30~ms) [Fischer et al., 2006a]. SEDs are most probably composed of several strokes, just like an Earth lightning flash, with stroke durations on the order of 100~$\mu$s [Mylostna et al., 2013], however, since there is no measured power spectral density for SED strokes, we do not consider the fine structure of lightning flashes in our study. The spectral roll-off, $n$, of an SED is not known, therefore we follow the approach of Farrell et al. [2007], and apply a slightly gentler roll-off, $n$ = 3.5, than the average Earth-value ($n = 4$). Substituting these values into Eq. \ref{eq:1}, we obtain the radiated power spectral density at the source of a single lightning flash at $f = 150$~MHz, to be $P/\Delta f = 3.8 \times 10^{-3}$~W~Hz$^{-1}$. 

The next step is to calculate the radio flux, of a single lightning flash, using Eq.~\ref{eq:2b}. The only unknowns are the power spectral density, which we determined to be $P/\Delta f = 3.8 \times 10^{-3}$~W~Hz$^{-1}$, and the distance of HAT-P-11, which is $d = 38$~pc,  The radio flux of one lightning flash under the assumptions listed above, therefore, is $I_{\mathrm{\nu, st}} = 2.2 \times 10^{-14}$~Jy. 

Finally, from Eq.~\ref{eq:2}, we express $n_\mathrm{tot,fl}$, the total number of lightning flashes necessary to produce an observed radio flux of $I_{\mathrm{\nu, obs}}$ = 3.87~mJy [L13]. We apply an average flash duration, $\tau_{\mathrm{fl}} = 0.3$~s, which is the largest event duration on Saturn according to Zarka et al. [2004], and an observation time of $\tau_{\mathrm{obs}} = 36$~min, which is the integration time for a single data point in L13 (their Fig. 2). The obtained value is $n_{\mathrm{tot,fl}} \approx 1.3 \times 10^{15}$ flashes. To compare our results with literature values of known lightning activity in the solar system, we convert $n_{\mathrm{tot,fl}}$ to flash density, $\rho_{\mathrm{fl}}$ applying Eq.~\ref{eq:2a}. With $\tau_{\mathrm{obs}} = 36$~min and $R_p \approx 0.4$ R$_{\mathrm{J}}$ (R$_{\mathrm{J}}$: Jupiter radius) [Bakos et al., 2010], we find that a thunderstorm with an average $\rho_{\mathrm{fl}} \approx 3.8 \times 10^5$ flashes km$^{-2}$ h$^{-1}$ is needed to produce the observed 3.87~mJy radio flux from the distance of HAT-P-11b, with parameters $n = 3.5$ and $\tau_\mathrm{fl} = 0.3$~s. 

\begin{figure}[ht]
\centering
\includegraphics[width=0.8\textwidth]{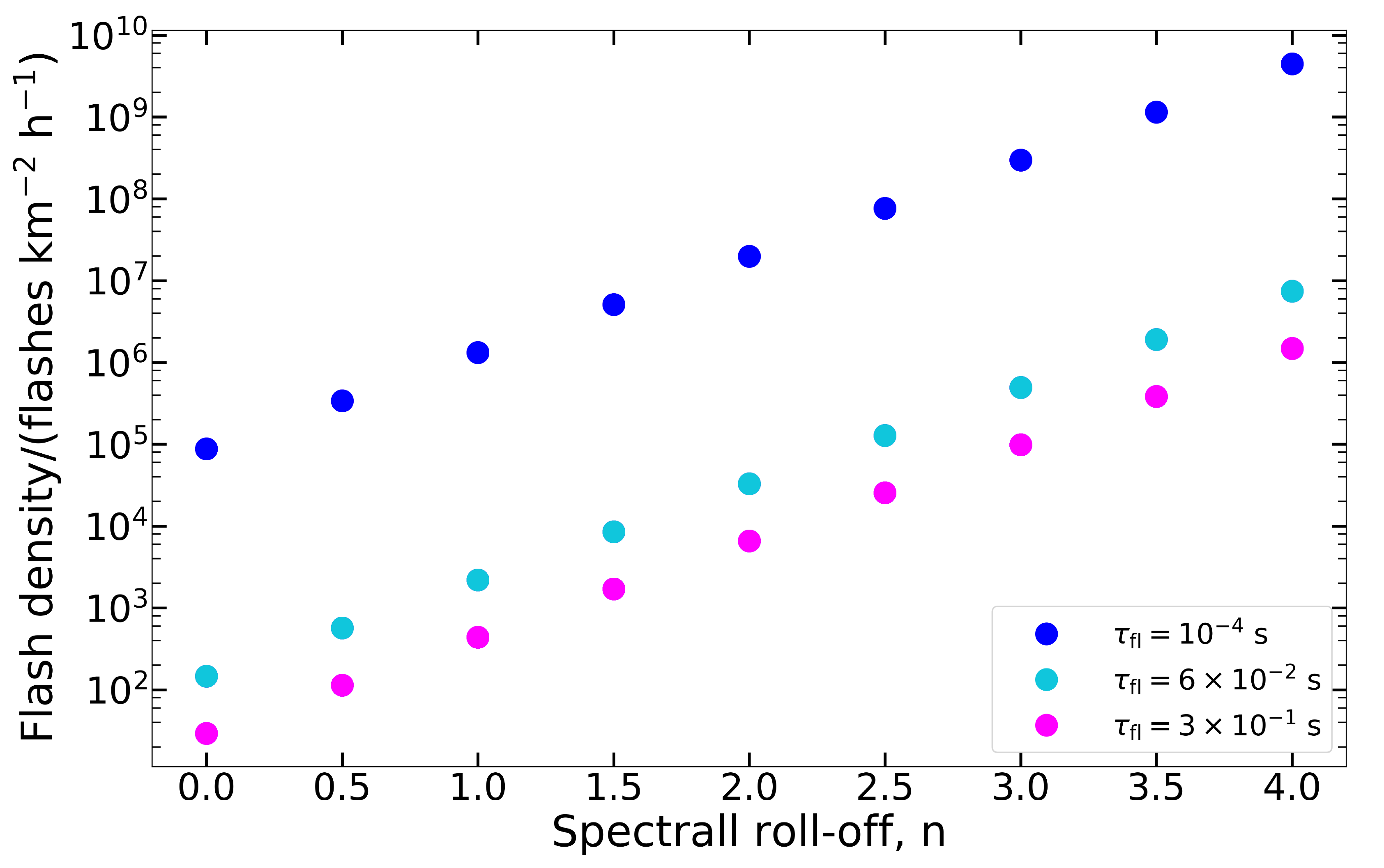}
\caption{Lightning flash density [flashes km$^{-2}$ h$^{-1}$], $\rho_{\mathrm{fl}}$ needed to emit the 3.87~mJy radio flux measured by Lecavelier des Etangs et al. [2013], depending on what spectral roll-off, $n$, is used in Eq. \ref{eq:1}. The figure also shows the dependence of $\rho_{\mathrm{fl}}$ on the flash duration, $\tau_{\mathrm{fl}}$ (Eq. \ref{eq:2}). For all the calculations, we assumed that the radiated power spectral density of lightning on HAT-P-11b is $P/\Delta f = 50$~W~Hz$^{-1}$ at $f = 10$~MHz, as was measured on Saturn [Fischer et al., 2006b]. The results show that the flatter the radiated power spectrum (the smaller $n$), the fewer flashes are needed to produce the observed radio flux ($\rho_{\mathrm{fl}}$ is smaller). For the same spectral roll-off, $n$, the slower discharges produce larger amount of power than quicker discharges.}
\label{fig:param}
\end{figure}

The spectral roll-off, $n$, and the flash duration, $\tau_\mathrm{fl}$, are not universal constants, there have been several different values observed for lightning in the solar system. Therefore, in this new study, we explore how various $n$ and $\tau_\mathrm{fl}$ parameters affect the obtained flash density, $\rho_{\mathrm{fl}}$, in the following parameter study. Zarka and Pedersen [1983] found the SED spectrum to be relatively flat ($n\approx 0.0$) between 1.2 and 40.2 MHz. Fischer et al. [2006b] found the roll-off of the SED spectrum to be $n=0.5$, between 4 and 16 MHz. The reason for such flat spectral behaviour of SEDs, however, is still unknown [G. Fischer, private communication]. The largest roll-off for lightning on Earth was found to be $n=4$ [Rakov and Uman, 2003]. Therefore, we conduct our study for $n = 0\,\ldots\,4$ with increments of 0.5. For each $n$, we calculate the power spectral density (Eq. \ref{eq:1}), the radio flux of a lightning flash (Eq. \ref{eq:2b}), and the flash density (Eq. \ref{eq:2a}). In each case, we apply a flash duration of $\tau_\mathrm{fl} = 10^{-4}, 6 \times 10^{-2}$, and $3 \times 10^{-1}$~s. 100~$\mu$s is the average stroke duration on Earth [Volland, 1984], and most probably on Saturn [Mylostna et al., 2013]. 60~ms is the average flash duration on Saturn [Zarka et al., 2006], while~300 ms is the largest flash duration on Saturn [Zarka et al., 2004]. We note that the duration of a lightning flash on an extrasolar object is unknown, therefore we use various values from the solar system, which were obtained for strokes, SEDs and flashes. The results of the parameter study are shown in Fig.~\ref{fig:param}. We find that the flatter the radiated power spectrum (the smaller $n$), the fewer flashes are needed to produce the observed radio flux ($\rho_{\mathrm{fl}}$ is smaller). Smaller $n$ means that the radiated radio power decreases much slower, therefore, one lightning flash will produce larger radio powers at higher frequencies, than what it would produce if the spectrum was steeper ($n$ was larger). For the same spectral roll-off, slower discharges release more power than quicker ones. In a best case, physical scenario with $n = 0.5$ and $\tau_\mathrm{fl} = 0.3$~s, a flash density $\rho_{\mathrm{fl}} = 114$ flashes km$^{-2}$ h$^{-1}$ would be enough to produce the observed radio flux of 3.87~mJy. We do not consider the $n=0$ values, as this would mean an overall flat spectrum of lightning, and an infinite amount of released power, which is unphysical. The most realistic case based on SED measurements ($n \simeq 0.5$, with $\tau_{\mathrm{fl}} = 60$~ms) would result in the requirement of $\rho_{\mathrm{fl,3}} \simeq 570$ flashes km$^{-2}$ h$^{-1}$. Throughout the paper, when we refer to the "realistic case", we consider the choice of $\tau_{\mathrm{fl}}$ and $n$ to be realistic and not the resulting flash density itself.
 
\begin{table}[h]
\caption{Lightning flash densities from HAT-P-11b (from this work and Hodos\'an et al., [2016a]) and examples from across the solar system. All values are from Hodos\'an et al., [2016b], apart from (*), which was obtained from [Huffines and Orville, 1999]. The top table lists values according to planets, the bottom one shows volcanic eruptions on Earth.}\label{table:1}
\begin{center}
\footnotesize
\begin{tabular}{llll}
	\hline 
	Planet & \vtop{\hbox{\strut Flash density}\hbox{\strut [km$^{-2}$ h$^{-1}$]}} & \vtop{\hbox{\strut Planetary hemisphere}\hbox{\strut surface area [km$^2$]}} & Comment \\
	\hline
	\textbf{HAT-P-11b} & $\mathbf{114}$ & $5.47 \times 10^9$ & $\rho_{\mathrm{fl,1}}$, $n=0.5$, $\tau_\mathrm{fl} = 0.3$; This work \\
	 & $\mathbf{3.8 \times 10^5}$ & & $\rho_{\mathrm{fl,2}}$, $n=3.5$, $\tau_\mathrm{fl} = 0.3$; This work \\
	 & $\mathbf{5.7 \times 10^2}$ & & $\rho_{\mathrm{fl,3}}$, $n=0.5$, $\tau_\mathrm{fl} = 0.06$; This work \\	
	Earth & 0.1 & $2.55 \times 10^8$ & largest average in the USA$^{(*)}$ \\
	 & $2.29 \times 10^{-4}$ & & global average from LIS/OTD data \\
	Saturn & $5.09 \times 10^{-6}$ & $2.13 \times 10^{10}$ & from Cassini (2010/11) data \\
	Jupiter & $1.43 \times 10^{-6}$ & $3.07 \times 10^{10}$ & from New Horizons (2007) data \\
	\hline
\end{tabular}

\vspace{0.5cm}

\begin{tabular}{lll}
	\hline 
	Volcano & Flash density [km$^{-2}$ h$^{-1}$] & Eruption \\
	\hline
	Eyjafjallaj\"okull & 0.1 & 2010 Apr 14$-$19 \\
	 & 0.32 & 2010 May 11$-$20 \\
	Mt Redoubt & 12.04 & 2009 Mar 23 \\
	 & $2 \times 10^3$ & 2009 Mar 29 (Phase 1) \\
	\hline
\end{tabular}
\end{center}
\end{table}

Table \ref{table:1} lists our result for HAT-P-11b in comparison to a few examples of flash densities observed in the solar system. We only list the best case ($\rho_{\mathrm{fl,1}} = 114$ flashes km$^{-2}$ h$^{-1}$), the example case ($\rho_{\mathrm{fl,2}} = 3.8 \times 10^5$ flashes km$^{-2}$ h$^{-1}$), and the most realistic case ($\rho_{\mathrm{fl,3}} = 570$ flashes km$^{-2}$ h$^{-1}$) for HAT-P-11b. This comparison shows that, a thunderstorm with a lightning occurrence rate obtained for the hypothetical storm on HAT-P-11b ranges between thunderstorms with flash densities of the same order of magnitude as the Mt. Redoubt eruption showed in 2009 March 23, and thunderstorms never seen in the solar system before. However, we have to remind ourselves, that the Jovian and Saturnian values are based on data from spacecraft, which can only observe the most energetic flashes from the planet [Hodos\'an et al., 2016b]. Secondly, we assumed that lightning on HAT-P-11b produces the same amount of energy and radio power that is known from the solar system. Hodos\'an et al. [2017, in prep.] showed that lightning can be more energetic and produce 2$-$8 orders of magnitude more radio power on hot exoplanets ($T_\mathrm{eff} = 1500 \dots 2000$~K) and brown dwarfs, than lightning on Earth. They used a dipole radiation model to investigate the effects of several modelling and physical properties of lightning (e.g. discharge duration, extension of the discharge, number of charges in the channel) on the energy and power release in radio wavelengths. The model was based on electric field and current models developed for Earth lightning [e.g. Bruce and Golde, 1941]. HAT-P-11b orbits the host star on a very close orbit, resulting in high atmospheric temperatures, and a planetary object that the solar system does not contain. Therefore, it is reasonable to think that if lightning exists on HAT-P-11b, it is more energetic and more frequent than lightning in the solar system.

\section{Conclusions} \label{sec:conc}
Here (also in Hodos\'an et al. [2016a]), we determined the lightning occurrence rate (flash density, $\rho_{\mathrm{fl}}$) of a hypothetical thunderstorm on the mini--Neptune HAT-P-11b, that could produce the average 3.87~mJy radio flux observed from the direction of the system by L13. We conducted a parameter study to investigate how the spectral roll-off, $n$, of the lightning radio spectrum, and the flash duration, $\tau_\mathrm{fl}$, affect the resulting $\rho_{\mathrm{fl}}$. 

Our results indicate that to produce a 3.87~mJy radio flux at 150~MHz, from the distance of HAT-P-11b, a thunderstorm with average flash densities many orders of magnitude larger than what has been observed in the solar system is required. However, in a best-case scenario, the flash density of the hypothetical storm on HAT-P-11b, is of the same order of magnitude as was observed during the 2009 March eruption of Mt Redoubt (Table \ref{table:1}). HAT-P-11b is a hot planet, larger than Earth, orbiting close to the host star, a type not known from the solar system. The close orbit may enhance lightning activity. Also, Hodos\'an et al. [2017, in prep.] showed that hot giant planets could produce lightning flashes that release a radio power up to 8 orders of magnitude larger than lightning on Earth, which would further support a "small flash density $-$ large radiated lightning power" scenario. However, we have to remember that in our study we assumed that the hypothetical thunderstorm covered the whole observable face of the planet. In the solar system, the most energetic lightning flashes were observed from Saturn, but the the occurrence rate of Saturnian lightning is limited only to a few percent of the total surface of the planet and in time they occur very rarely [e.g. summary in Hodos\'an et al. 2016b]. Similarly, even though the flash density of volcanic eruptions is very large on Earth, it is limited to a short period of time and a small area on the planet [e.g. James et al., 2008].

We conclude that lightning on exoplanets may produce observable radio signatures if its radio spectrum around 10~MHz is similar to Saturnian lightning spectra, and above that it becomes close to flat. In this case, from the distance of HAT-P-11b, lightning storms with flash densities ten times bigger than the average largest ones in the USA, would be enough to produce radio emission with a flux of a few mJy, if the thunderstorm covers half the planet. However, considerations regarding occurrence rates of lightning on Saturn and Earth, and our results for the most realistic Saturnian-like flash density case ($n = 0.5$, $\tau_{\mathrm{fl}} = 0.06$~s) suggest that the 3.87 mJy radio emission observed from the direction of HAT-P-11b was not produced by lightning activity, since the required flash density of $\rho_{\mathrm{fl,3}} = 570$ flashes km$^{-2}$ h$^{-1}$ is unrealistically high. For future lightning observations a combined radio and optical observing campaign is desirable, especially when directly imaged planets and close brown dwarfs are studied. Such observations should be followed-up by infrared telescopes looking for signatures of chemical changes in the atmosphere, caused by lightning activity, as was suggested in Hodos\'an et al. [2016a].

\textbf{Acknowledgements.} We thank Zach Cano, Brice-Olivier Demory, Aurora Sicilia-Aguilar, Alain Lecavelier des Etangs, Philippe Zarka, and Georg Fischer for useful discussions. We highlight financial support of the European Community under the FP7 by an ERC starting grant number 257431. ChH highlights the hospitality of the Universiteit van Amsterdam, and travel support from NWO and LKBF.

\section*{References}
\everypar={\hangindent=1truecm \hangafter=1}


Ardaseva,~A., P.\,B.~Rimmer; I.~Waldmann, M.~Rocchetto, S.\,N.~Yurchenko, Ch.~Helling, J.~Tennyson, Lightning chemistry on Earth-like exoplanets, \textsl{MNRAS}, \textbf{470}, 187--196, 2017

Bakos,~G.\,\'A., G.~Torres, A.~P{\'a}l, J.~Hartman, G.~Kov{\'a}cs, R.\,W.~Noyes, D.\,W.~Latham, D.\,D.~Sasselov, B.~Sip{\H o}cz, G.\,A.~Esquerdo et al., HAT-P-11b: A super--Neptune planet transiting a bright K star in the Kepler field, \textsl{ApJ}, \textbf{710}, 1724--1745, 2010.

Brook,~M., M.~Nakano, P.~Krehbiel, and T.~Takeuti, The electrical structure of the Hokuriku winter thunderstorms, \textsl{J. Geophys. Res.}, \textbf{87}, 1207--1215, 1982.

Bruce,~C.\,E.\,R., and R.\,H.~Golde, The lightning discharge. \textsl{The Journal of the Institution of Electrical Engineers}, \textbf{88}, 487--520, 1941

Dyudina,~U.\,A., A.\,D.~Del Genio, A.\,P.~Ingersoll, C.\,C.~Porco, R.\,A.~West, A.\,R.~Vasavada, and J.\,M.~Barbara, Lightning on Jupiter observed in the H$_{\alpha}$ line by the Cassini imaging science subsystem, \textsl{Icarus}, \textbf{172}, 24--36, 2004.

Dyudina,~U.\,A., A.\,P.~Ingersoll, S.\,P.~Ewald, C.\,C.~Porco, G.~Fischer, and Y.~Yair, Saturn's visible lightning, its radio emissions, and the structure of the 2009--2011 lightning storms, \textsl{Icarus}, \textbf{226}, 1020--1037, 2013.

Farrell,~W.\,M., M.\,L.~Kaiser, G.~Fischer, P.~Zarka, W.\,S.~Kurth, and D.\,A.~Gurnett, Are Saturn electrostatic discharges really superbolts? A temporal dilemma, \textsl{Geophys. Res. Lett.}, \textbf{34}, L06202, 2007.

Fischer,~G., M.\,D.~Desch, P.~Zarka, M.\,L.~Kaiser, D.\,A.~Gurnett, W.\,S.~Kurth, W.~Macher, H.\,O.~Rucker, A.~Lecacheux, W.\,M.~Farrell, and B.~Cecconi, Saturn lightning recorded by Cassini/RPWS in 2004,
\textsl{Icarus}, \textbf{183}, 135--152, 2006a.

Fischer,~G., W.~Macher, M.\,D.~Desch, M.\,L.~Kaiser, P.~Zarka, W.\,S.~Kurth, W.\,M.~Farrell, A.~Lecacheux, B.~Cecconi, and D.\,A.~Gurnett, On the intensity of Saturn lightning, in \textsl{Planetary Radio Emissions VI}, edited by H.\,O.~Rucker, W.\,S.~Kurth, and G.~Mann, Austrian Academy of Sciences Press, Vienna, 123--132, 2006b.

Fraine,~J., D.~Deming, B.~Benneke, H.~Knutson, A.~Jord{\'a}n, N.~Espinoza, N.~Madhusudhan, A.~Wilkins, and K.~Todorov, Water vapour absorption in the clear atmosphere of a Neptune--sized exoplanet, \textsl{Nature}, \textbf{513}, 526--529, 2014.

Grie{\ss}meier,~J.--M., P.~Zarka, and H.~Spreeuw, Predicting low-frequency radio fluxes of known extrasolar planets, \textsl{Astron. Astrophys.}, \textbf{475}, 359--368, 2007.

Gurnett,~D.\,A., W.\,S.~Kurth, I.\,H.~Cairns, and L.\,J.~Granroth, Whistlers in Neptune's magnetosphere: Evidence for atmospheric lightning, \textsl{JGR}, \textbf{95}, 20967--20976, 1990.	

Helling,~Ch., M.~Jardine, D.~Diver, and S.~Witte, Dust cloud lightning in extraterrestrial atmosphere, \textsl{Planet. Space Sci.}, \textbf{77}, 152--157, 2013.

Helling,~Ch.,	R.\,G.~Harrison, F.~Honary, D.\,A.~Diver, K.~Aplin, I.~Dobbs-Dixon, U.~Ebert, S.~Inutsuka, F.\,J.~Gordillo-Vazquez, and S.~Littlefair, 
	Atmospheric Electrification in Dusty, Reactive Gases in the Solar System and Beyond, \textsl{Surv. in Geophys.}, \textbf{37}, 705--756, 2016a.

Helling,~Ch., G.~Lee, I.~Dobbs--Dixon, N.~Mayne, D.\,S.~Amundsen, J.~Khaimova, A.\,A.~Unger, J.~Manners, D.~Acreman, and C.~Smith, The mineral clouds on HD 209458b and HD 189733b, \textsl{MNRAS}, \textbf{460}, 855--883, 2016b.

Hodos\'an,~G., P.\,B.~Rimmer, and Ch.~Helling, Is lightning a possible source of the radio emission on HAT-P-11b? \textsl{MNRAS}, \textbf{461}, 1222--1226, 2016a.

Hodos\'an,~G., Ch.~Helling, R.~Asensio--Torres, I.~Vorgul, and P.\,B.~Rimmer, Lightning climatology of exoplanets and brown dwarfs guided by solar system data, \textsl{MNRAS}, \textbf{461}, 3927--3947, 2016b.

James,~M.\,R., L.~Wilson, S.\,J.~Lane, J.\,S.~Gilbert, T.\,A.~Mather, R.\,G.~Harrison, and  R.\,S.~Martin, Electrical charging of volcanic plumes, \textsl{Space Sci. Rev.}, \textbf{137}, 399--418, 2008.

Kasting,~J.\,F., Earth's early atmosphere, \textsl{Science}, \textbf{259}, 920--926, 1993.

Lecavelier~des~Etangs,~A., S.\,K.~Sirothia, Gopal--Krishna, and P. Zarka, Hint of 150~MHz radio emission from the Neptune--mass extrasolar transiting planet HAT-P-11b, \textsl{Astron. Astrophys.}, \textbf{552}, A65, 2013.

Line,~M.\,R. and V.~Parmentier, The influence of nonuniform cloud cover on transit transmission spectra, \textsl{ApJ}, \textbf{820}, 78, 2016.

Longstaff,~E.\,S., S.\,L.~Casewell, G.\,A.~Wynn, P.\,F.\,L.~Maxted, and Ch.~Helling, Emission lines in the atmosphere of the irradiated brown dwarf WD0137-349B, \textsl{MNRAS}, \textbf{471}, 1728--1736, 2017

Lopez,~E.\,D. and J.\,J.~Fortney, Understanding the mass--radius relation for sub--neptunes: Radius as a proxy for composition, \textsl{ApJ}, \textbf{792}, 1, 2014.

Mayor,~M. and D.~Queloz, A Jupiter--mass companion to a solar-type star, \textsl{Nature}, \textbf{378}, 355--359, 1995.

Miller,~S.\,L. and H.\,C.~Urey, Organic compound synthesis on the primitive Earth, \textsl{Science}, \textbf{130}, 245--251, 1959.

Mylostna,~K.\,Y., V.\,V. Zakharenko, A.\,A.~Konovalenko, V.~Kolyadin, P.~Zarka, J.--M.~Grie{\ss}meier, G.~Litvinenko, M.~Sidorchuk, H.\,O.~Rucker, G.~Fischer, B.~Cecconi, A.~Coffre, L.~Denis, V.~Nikolaenko, and V. Shevchenko, Study of Saturn Electrostatic Discharges in a wide range of time scales, \textsl{Odessa Astronom. Pub.}, \textbf{26}, 251, 2013.

Patel,~B.\,H., C.~Percivalle, D.\,J.~Ritson, C.\,D.~Duffy, and J.\,D.~Sutherland, Common origins of RNA, protein and lipid precursors in a cyanosulfidic protometabolism, \textsl{Nature Chemistry}, \textbf{7}, 301--307, 2015.

Rakov,~V.\,A., and M.\,A.~Uman, Lightning: Physics and Effects, Cambridge Univ. Press, Cambridge, UK, 2003.

Rinnert,~K., L.\,J.~Lanzerotti, M.\,A.~Uman, G.~Dehmel, F.\,O.~Gliem, E.\,P.~Krider, and J.~Bach, Measurements of radio frequency signals from lightning in Jupiter's atmosphere, \textsl{J. Geophys. Res.}, \textbf{103}, 22979--2299, 1998.

Russell,~C.\,T., Planetary magnetospheres, \textsl{Rep. Prog. Phys.}, \textbf{56}, 687--732, 1993.

Russell,~C.\,T., T.\,L.~Zhang and H.\,Y.~Wei, Whistler mode waves from lightning on Venus: Magnetic control of ionospheric access, \textsl{J. Geophys. Res.}, \textbf{113}, E00B05, 2008.

Volland,~H., Atmospheric electrodynamics, \textsl{Springer--Verlag}, 1984.

Williams,~P.\,K.\,G. and E.~Berger, The rotation period and magnetic field of the T dwarf 2MASSI J1047539+212423 measured from periodic radio bursts, \textsl{ApJ}, \textbf{808}, 189, 2015.

Wilson,~C.\,T.\,R., Investigations on lightning discharges and on the electric field of thunderstorms, \textsl{Phil. Trans. Royal Soc. London S. A}, \textbf{221}, 73--115, 1921.	

Yair,~Y., New results on planetary lightning, \textsl{Adv. Space Res.}, \textbf{50}, 293--310, 2012.

Yair,~Y., S.~Katz, R.~Yaniv, B.~Ziv and	C.~Price, An electrified dust storm over the Negev desert, Israel, \textsl{Atmos. Res.}, \textbf{181}, 63--71, 2016.	

Zarka,~P., and B.\,M.~Pedersen, Statistical study of Saturn electrostatic discharges, \textsl{J. Geophys. Res.}, \textbf{88}, 9007--9018, 1983.

Zarka,~P., and B.\,M.~Pedersen, Radio detection of Uranian lightning by Voyager~2, \textsl{Nature}, \textbf{323}, 605--608, 1986.

Zarka,~P., W.\,M.~Farrell, M.\,L.~Kaiser, E.~Blanc, and W.\,S.~Kurth, Study of solar system planetary lightning with LOFAR, \textsl{Planet. Space Sci.}, \textbf{52}, 1435--1447, 2004.

Zarka,~P., B.~Cecconi, L.~Denis, W.\,M.~Farrell, G.~Fischer, G.\,B.~Hospodarsky, M.\,L.~Kaiser, and W.\,S.~Kurth, Physical properties and detection of Saturn's lightning radio bursts, in \textsl{Planetary Radio Emissions VI}, edited by H.\,O.~Rucker, W.\,S.~Kurth, and G.~Mann, Austrian Academy of Sciences Press, Vienna, 111--122, 2006.

\end{document}